\documentclass[notoc]{JHEP} 
\usepackage{amsmath}
\usepackage{amssymb,amsfonts}
\newcommand{\mide}{{\rm mid}}

\newcommand{\Tr}{{\rm Tr}\ }
\newcommand{\A}{{\cal A}}
\newcommand{\B}{{\cal B}}
\newcommand{\C}{{\cal C}}
\newcommand{\D}{{\cal D}}

\newcommand{\K}{{\cal K}}

\newcommand{\Zint}{\mathbb{Z}}
\newcommand{\Real}{\mathbb{R}}

\title{Morita equivalence and T-duality\\(or $B$ versus $\Theta$)}

\author{B. Pioline\thanks{Work supported in part by TMR networks
ERBFMRXCT96-0045 and ERBFMRXCT96-0090.}\\
Centre de Physique Th{\'e}orique\thanks{Unit{\'e} mixte
CNRS UMR 7644}, \\ Ecole
Polytechnique, {}F-91128 Palaiseau, France\\
\email{pioline\@cpht.polytechnique.fr}}

\author{A. Schwarz\\Institut des Hautes Etudes Scientifiques,\\
Le Bois-Marie, {}F-91440 Bures-sur-Yvette, France\\[1mm]
Dept. of Mathematics, University of California,\\
Davis, CA 95616 USA\\
\email{schwarz@math.ucdavis.edu}}

\abstract{T-duality in M(atrix) theory has been argued to be realized as 
Morita equivalence in Yang-Mills theory on a non-commutative 
torus (NCSYM). Even though the two have the same structure group,
they differ in their action since Morita
equivalence makes crucial use of an additional modulus
on the NCSYM side, the constant Abelian magnetic background.
In this paper, we reanalyze and clarify the correspondence between
M(atrix) theory and NCSYM, and provide two resolutions of this puzzle.
In the first of them, the standard map is kept and the extra modulus
is ignored, but the anomalous transformation is offset by the
M(atrix) theory ``rest term''. In the second, the standard map
is modified so that the duality transformations agree, and 
a $SO(d)$ symmetry is found to eliminate the spurious modulus.
We argue that this is a true symmetry of supersymmetric Born-Infeld 
theory on a non-commutative torus, which allows to freely trade
a constant magnetic background for non-commutativity
of the base-space. We also obtain a BPS mass formula for this theory, invariant
under T-duality, U-duality, and continuous $SO(d)$ symmetry.}

\preprint{CPHT--S730-0899\\IHES/P/99/64\\hep-th/9908019}
\keywords{M(atrix) Theories, D-branes, M-Theory, String Duality}

\begin{document}

Gauge theories on non-commutative spaces arise 
in M(atrix) theory compactifications with non-vanishing three-form
$\C_{ij-}$ on the light-cone \cite{Connes:1998cr}, and
in string theory as world-volume theories of D-branes in 
large B-field
\cite{Douglas:1998fm}. This effect is most
easily understood in toroidal compactifications after T-duality: 
the torus with large B-field turns into a slanted torus,
and the non-locality characteristic of non-commutative geometry
arises from the shortest open strings relating the D-brane
to its translated images as the torus becomes more slanted 
\cite{Douglas:1998fm,Li:1998ks}. In fact, in the presence
of a non-vanishing B-field, -- or $U(1)$ field strength,
since only the combination $F+B$ is gauge-invariant --
the longitudinal coordinates at the end of the open strings 
attached to the D-brane become non-commutative 
\cite{Chu:1998qz,Ardalan:1999av,Schomerus:1999ug}, 
revealing that the world-volume itself has turned into
a non-commutative space. Therefore, the worldvolume theory of 
D-branes should described by Born-Infeld theory
on a non-commutative space \cite{Hofman:1998iy}.
Unfortunately, this theory lacks a precise definition,
for the same reason as non-Abelian Born-Infeld theory does.
In this work, we shall follow a rather indirect route
to uncover some of its properties.

The dynamics of $N$ D0-branes on $T^d$ in the Sen-Seiberg limit of
vanishing volume \cite{Sen:1997we,Seiberg:1997ad}
is supposed to describe M-theory on the light-cone
\cite{Banks:1997vh,Susskind:1997cw}
and the background B-field, hold fixed in the limit, simply corresponds 
to the vev of the three-form
field strength $C_{ij-}$ on the lightlike direction \cite{Connes:1998cr}. 
The BPS mass formula on the M-theory side is known from 
U-duality considerations \cite{Obers:1997kk} 
(see \cite{Obers:1998fb,Obers:1998rn} 
for a review), whereas the BPS energy spectrum in the non-commutative 
super-Yang-Mills theory (NCSYM) was derived rigorously in 
\cite{Konechny:1999rz}. This theory is invariant under a
$SO(d,d,\Zint)$ group generated by Morita equivalence,
which was argued to reflect the T-duality invariance
of M-theory on a torus $T^{d+1}$ \cite{Connes:1998cr,Schwarz:1998qj}.
Further studies have followed in \cite{Ho:1998hq,Brace:1998ku,
Konechny:1998wv,Brace:1998xz}. The BPS spectrum even organizes itself
in multiplets of the extended U-duality group $E_{d+1(d+1)}(\Zint)$,
although only T-duality is a symmetry of the mass formula
\cite{Hacquebord:1997nq,Blau:1997du,Obers:1997kk,Hull:1997jc}.
The correspondence between the M-theory and gauge theory sides
nevertheless encounters several difficulties. For one thing, the NCSYM theory
seems to offer an extra modulus $\Phi_{ij}$, which corresponds
to a constant Abelian magnetic background, and for another,
T-duality and Morita equivalence 
act differently on the two sides, even
though representing the action of the same $SO(d,d,\Zint)$ group.
Our first aim in this paper is to reconcile the two sides
and clarify this correspondence. 

Our first approach takes advantage of the fact that the
Abelian magnetic background couples to the Yang-Mills fields
through topological terms, and thus can be set to zero
upon proper account of the induced spectral flow. The resulting
theory is no longer invariant under Morita equivalence, 
but we shall argue that the anomalous variation is offset
by a ``rest term'' which is naturally motivated from the M(atrix) 
side. In this way, we recover the standard mapping between
M-theory and NCSYM, and confirm the identification of
the M-theory B-field with the NCSYM deformation parameter $\Theta$.
We also obtain a simpler relation between
NCSYM energies and M-theory masses than usually presented.
This will be discussed in section 1.

In our second approach, we pay full respect to the dual parameters
$\Theta$ and $\Phi$ of the NCSYM, and exhibit a map to the M-theory
B-field modulus which translates Morita equivalence
into the standard T-duality. This map is not one-to-one, and 
indeed we uncover a continuous $SO(d)$ symmetry relating the various
preimages of a given $B$. These map and symmetry can be found in
equations (\ref{map},\ref{so2}) in section 2 of the present paper for the
$d=2$ case, and (\ref{mapd},\ref{sod}) for the general case.
We propose that this invariance is a
true symmetry of the putative Born-Infeld theory on a non-commutative
torus, and exhibit an T-duality and U-duality invariant mass formula 
(\ref{genmassbi})
which reduces to the NCSYM mass formula in the  ``non-relativistic'' 
$l_s\to 0$ limit (in the sense of \cite{Townsend}), and to the usual
Born-Infeld mass formula for a commutative torus.
Our result implies that a magnetic background $\Phi$ can be freely traded
for a non-commutativity $\Theta$ of the base-space, through
a $SO(d)$ transformation.  This equivalence between base-space 
non-commutativity and magnetic
background fits nicely with the observations reported in \cite{NCSW}.
Unfortunately, for lack of an
appropriate formulation of non-commutative Born-Infeld theory,
proving that this invariance is a full-fledged symmetry
will remain beyond reach.

\section{M-theory and M(atrix) gauge theory}

\noindent{\it M-theory mass spectrum and T-duality.}
The BPS mass formula for M-theory compactified on a torus $T^d$ times
a circle of radius $R_s$ in the presence of general gauge backgrounds
was obtained in \cite{Obers:1997kk} from U-duality requirements. 
For simplicity, we shall at first focus
on the $d=2$ case, where it shrinks to a more manageable form,
\begin{align}
\label{omass}
{\cal M}^2=&\frac{1}{R_s^2} (m_s+B_{ij} m^{ij})^2 +
\left[ (m_i+B_{ij} m^{js}) + A_i (m_s+B_{jk} m^{jk} )\right]^2\nonumber\\
&+\frac{(m^{ij})^2}{l_M^6} 
+\frac{R_s^2}{l_M^6}\left( m^{is}+ A_j m^{ij} \right)^2
+\sqrt{\frac{(K^i)^2+R_s^2 (K^s+A_i K^i)^2}{l_M^6}}
\end{align}
Here, $l_M$ denotes the 11D Planck length, $A_i$ and $B_{ij}$ 
the off-diagonal metric $G_{is}$ and
three-form $C_{ijs}$ where $s$ stands for the direction of radius $R_s$,
and the contractions are performed with respect to the metric $G_{ij}$
or inverse metric $G^{ij}$ on the torus $T^d$. The integers $m_s,m_i$,
$m^{is},m^{ij}$ and $K^i,K^s$ stem from the reduction of the charges
$m_I,m^{IJ}$ and $K^I$ of the particle and string multiplets on 
$T^2\times S^1$, respectively. In particular, the string charges
are related to the particle charges through $K^I=m_J m^{IJ}$
for 1/4-BPS states, which
has to vanish for 1/2-BPS states. Their physical interpretation 
is clear from their contributions
to the mass: $m_s$ and $m_i$ are the momenta around $S_1$ and $T^2$,
$m^{is}$ and $m^{ij}$ are the membrane charge on $S^1\times T^2$.
This mass formula is invariant under the U-duality group $Sl(3,\Zint)
\times Sl(2,\Zint)$ on $T^2\times S^1=T^3$, under which the momentum
multiplet charges transforms as $(3,2)$ and the string charges as
$(3,1)$. We shall be particularly interested in the T-duality
subgroup $Sl(2,\Zint)_T\subset Sl(3,\Zint)$, whose action is
most easily seen after going to string theory variables
$l_s^2=l_M^3/R_s$, $g_s l_s = R_s$:
\begin{align}
\label{tmass}
\begin{split}
{\cal M}^2=&\frac{T_2}{g_s^2 l_s^2} \frac{|m-nT|^2}{T_2} 
+\sqrt{ \frac{T_2}{g_s^2 l_s^4} K^i \hat G_{ij} K^j + \frac{ K^2}{l_s^4} }\\
&+ \frac{1}{l_s^2 T_2} \left[(m_i+A_i m)-(n_i+A_i n)T\right] \hat G^{ij}
\left[(m_j+A_j m)-(n_j+A_j n)\bar T\right]  
\end{split}
\end{align}
where $T= T_1+i T_2 =B_{12}+i \sqrt{\det G_{ij}}/{l_s^2}$ is the
standard complex modulus and $\hat G_{ij}=G_{ij}/\sqrt{\det G}$ is 
the unit volume metric of the torus $T^2$. We also relabelled 
the charges as $m=m_s,n=\epsilon_{ij}m^{ij}/2,n_i=\epsilon_{ij}m^{js},
K=K^s=\epsilon^{ij}m_i n_j, K^i=\epsilon^{ij}(m n_i- n m_i)$.
The mass formula is then invariant under the standard $Sl(2,\Zint)_T$ 
T-duality
\begin{gather}
\label{tdualsta}
T\to \frac{aT+b}{cT+d}\ ,\quad 
g_s\to \frac{g_s}{|cT+d|}\ ,
\begin{pmatrix} m & m_i \\ n & n_i \end{pmatrix} \to 
\begin{pmatrix}a & b\\c & d \end{pmatrix}
\begin{pmatrix}m & m_i \\ n & n_i \end{pmatrix}\ ,\quad
\end{gather}
where $ad-bc=1$, leaving $\hat G_{ij}$ and $k_i$ invariant. Note 
for later reference the following useful consequences:
\begin{gather}
T_1\to \frac{a T_1 + b}{c T_1 + d} +
\frac{c T_2^2}{(cT_1 + d) |cT+d|^2}\ ,\quad
T_2\to \frac{T_2}{|cT + d|^2}\ ,\quad
m-nT\to\frac{m-nT}{cT_1+d}
\end{gather}

Following Sen and Seiberg \cite{Sen:1997we,Seiberg:1997ad}, 
the Discrete Light-Cone Quantization of
M-theory on a torus $T^d$ with light-cone momentum $m$ is equivalent to 
M-theory compactified on $T^d$ times a space-like circle of radius $R_s$ 
with momentum $m$, in the scaling limit $R_s\to 0$,
\begin{gather}
l_s = \frac{R_s^{1/2}}{M^{3/2}}\ ,\quad
g_s = (R_s M)^{3/4}\ ,\quad 
G_{ij} = \frac{R_s}{M} \gamma_{ij} \ ,\quad
\end{gather}
holding $M$ and $\gamma_{ij}$ fixed. Under this scaling, the mass formula
reduces to
\begin{align}
\label{mmass}
\begin{split}
{\cal M}=&\frac{|m-n B|}{g_s l_s} 
+\frac{T_2^2 n^2}{2 g_s l_s |m-nB|} \\
&+ \frac{g_s}{2 l_s T_2|m-nB|}
\left[(m_i+A_i m)-(n_i+A_i n)B\right] \hat G^{ij}\left[
(m_j+A_j m)-(n_j+A_j n)B\right]\\
&+\frac{\sqrt{T_2}}{l_s|m-nB|} \sqrt{K^i \hat G_{ij} K^j}
+{\cal O}(R_s) 
\end{split}
\end{align}
while the interactions between the D0-particles decouple
from the closed string bulk modes, and when $A=B=0$ 
reduce to a $U(m)$ super-Yang-Mills theory on the dual
torus. In this limit, the relativistic dispersion relation (\ref{tmass})
has turned into its non-relativistic Galilean limit (\ref{mmass}).

\vskip 3mm
\noindent{\it Non-commutative Yang-Mills and Morita equivalence.}
On the other hand, the BPS energy formula of 
super-Yang-Mills theory on a non-commutative torus $T^2_\theta$
given by the action functional \footnote{The trace $\Tr$ for a commutative
torus  reduces to the integral $\int d^dx /V$.}
\begin{equation}
\label{ncsyma}
S=-\frac{V}{4g_{YM}^2} \Tr (F_{ij} + \Phi_{ij}\cdot 1)^2
+i \frac{V}{2g_{YM}^2} \Tr \bar\psi\Gamma^i[\nabla_i,\psi]
+\lambda^i \Tr F_{0i}
\end{equation}
where $\Phi_{ij}=\phi\epsilon_{ij}$, 
was computed in \cite{Konechny:1999rz} from a study of the central charges
in the BPS algebra, with the result that
\begin{align}
\label{energy}
E=\frac{1}{2V g_{YM}^2 \dim E} (q + \phi \dim E)^2+
\frac{g_{YM}^2 }{2V \dim E} (n^i-\theta m^i + \lambda^i \dim E)^2 
+ \frac{1}{\dim E} \sqrt{(k_i)^2}
\end{align}
with $\dim E=p-q\theta$ and $k_i=\epsilon_{ij}(p m^i - q n^i)$.
Contractions are performed with the metric $g_{ij}$ or inverse metric
$g^{ij}$ depending on the position of the indices.
This formula is invariant under the $Sl(2,Z)_M$ action
\begin{subequations}
\label{tdualexo}
\begin{gather}
\theta\to \frac{a\theta+b}{c\theta+d}\ ,\quad
\phi\to(c\theta+ d)^2 \phi - c (c\theta+ d) \ ,\quad
g_{ij}\to (c\theta+d)^2g_{ij}\ ,\quad \\ 
g_{YM}^2\to (c\theta+d)g_{YM}^2\ ,\quad
\begin{pmatrix} p & n^i \\ q & m^i \end{pmatrix} \to 
\begin{pmatrix}a & b\\c & d \end{pmatrix}
\begin{pmatrix}p & n^i \\ q & m^i \end{pmatrix}\ ,\quad
\end{gather}
\end{subequations}
leaving $k_i$ invariant. It is useful to note that this transformation
implies
\begin{equation} V\to(c\theta+d)^2V\ ,\quad
\dim E \to \dim E/(c\theta+d)\ ,\quad
\mide E \to \mide E(c\theta+d)
\end{equation}
where we suggestively defined $\mide E=q + \phi \dim E$.
This invariance was deduced in \cite{Schwarz:1998qj} from Morita equivalence
of non-commutative tori $T_\theta$. It is important to notice
that this transformation involves the magnetic background field $\phi$,
as also recognized in \cite{Connes:1998cr,Ho:1998hq}.

\vskip 3mm
\noindent {\it Morita equivalence and T-duality reconciled.}
We now would like to reconcile the
M-theory mass formula (\ref{mmass}) and NCSYM mass formula (\ref{energy}).
The two certainly look very similar, but there are some important differences.
First of all, the first term in (\ref{mmass}), linear in the charges,
has no counterpart in (\ref{energy}): it may be included on the gauge
theory side by adding an innocuous term
\begin{equation}
S_{rest}= \frac{V}{g_{YM}^2 l_s^4} \Tr 1\ ,
\end{equation}
to the action functional (\ref{ncsyma}). Indeed, this term is simply
the constant term in the Born-Infeld action, and gives the mass of
the $m$ D$d$-branes on which the gauge theory lives. Such a linear
term was also introduced by hand in \cite{Connes:1998cr} to enforce
invariance. We will find more hints for the relevance of Born-Infeld
theory shortly.

The second major difference is in the
extra modulus $\phi$ which appears on NCSYM side and has no obvious
counterpart on the M-theory side. When $\phi=0$ and up to
the rest term, we obtain agreement under the map
\begin{subequations}
\begin{gather}
\label{mmap}
(p,q)=(m,n)\ ,\quad
(n^i,m^i)=(m_i,n_i)\ ,\quad k_i=K^i\ ,\\\quad g_{ij}=l_s^4 G^{ij}\ ,
\quad g_{YM}^2=\frac{g_s l_s^{2d-3}}{\sqrt{\det(G_{ij})}}\ ,\quad
\theta=B\ ,\quad \lambda^i=A_i 
\end{gather}
\end{subequations}
which implies $\hat g_{ij}=\hat G^{ij}, V=\frac{l_s^2}{T_2}$ and
reduces to the standard map at $B=\theta=0$.
However, we should still understand how the two $Sl(2,\Zint)$
transformation rules (\ref{tdualsta}) and (\ref{tdualexo})
relate to each other.

One resolution of the puzzle comes in the following way. First, one
should notice that although the modulus $\phi$ appears naturally
in the proof of $SO(d,d,\Zint)$ Morita invariance of NCSYM, one can always
exclude it from the final formula: its only contribution to the
action functional (\ref{ncsyma}) is through topological terms,
and consequently the energy for $\phi\neq 0$ can be deduced from
the one at $\phi=0$ through
\begin{equation}
E(\phi)=E(\phi=0)+\frac{V}{4g_{YM}^2} \left(
2 \Phi_{ij} \Tr F^{ij} + (\Phi_{ij})^2 \Tr 1 \right)
\end{equation}
We have $\Tr 1=\dim E=p-q\theta$, and the central charge $\Tr F_{ij}$ 
can be calculated in terms of topological
numbers by means of the formula relating the generating function of
topological numbers
$\mu(E)$ to the Chern character ${\rm ch(E)}$, which for $d=2$
implies that $\Tr F_{ij}=q \epsilon_{ij}$. 
In this way, we recover the formula (\ref{energy}). 
Using the above formula, we can thus easily derive the transformation rule 
for the energy at $\phi=0$:
\begin{equation}
\label{noninv}
E(\phi=0)\to E(\phi=0)+c\frac{2 (cp+dq)-c|p-q\theta|}
{2Vg_{YM}^2 (c\theta+d)}
\end{equation}
the rest term being strictly invariant.

Let us now consider the M-theory side.
The exact mass formula ${\cal M}^2$ is invariant under
the standard T-duality (\ref{tdualsta}). In the limit $R_s\to 0$ and
therefore $T_2\to 0$, the transformation rules (\ref{tdualsta})
reduce at leading order to the NCSYM ones (\ref{tdualexo}).
Under these truncated transformation rules, the rest term is
thus invariant, but the quadratic term changes as in (\ref{noninv}).
However, there are subleading corrections to the transformation rules,
\begin{subequations}
\begin{eqnarray}
T_1&\to& \frac{a T_1 + b}{c T_1 + d} + \frac{c T_2^2}{(c T_1+d)^3}+\dots \\
T_2&\to& \frac{T_2}{(cT_1 + d)^2} - \frac{c^2 T_2^3}{(cT_1+d)^4}+\dots
\end{eqnarray}
\end{subequations}
such that the rest term is no more invariant, but instead
varies in the opposite way as the quadratic term (\ref{noninv}),
thus cancelling the anomalous variation of the energy.
In this way, we have reconciled the two apparently different 
transformation rules (\ref{tdualsta}) and (\ref{tdualexo}).

\vskip 3mm
\noindent {\it M-theory and NCSYM on higher-dimensional tori.}
We now would like to show that this correspondence holds in more general
cases. The mass formula for M-theory on a torus $T^{d}$ times a circle
is known in full
generality from \cite{Obers:1998fb}. In order
to keep things manageable and given the difficulties in defining
the Matrix gauge theory on $T^d$, $d>4$, we shall restrict 
ourselves to $d\leq 4$. In addition to the charges appearing in (\ref{omass}),
we need to introduce those coming from the reduction of the
charges $m^{ijklm}$ and $K^{ijkl}$ 
of the particle (or flux) and string (or momentum)
multiplets on $T^{d+1}$. We should also take into account the vev $\C_3$ of the
M-theory three-form on the torus $T^d$. The BPS mass formula then reads
\begin{align}
\label{genmass}
{\cal M}^2=&\frac{1}{R_s^2} (\tilde m_s)^2 +(\tilde m_i)^2
+\frac{(m^{ij})^2}{2l_M^6} 
+\frac{R_s^2}{l_M^6}(\tilde m^{is})^2
+\frac{R_s^2}{4!~l_M^{12}} ( \tilde m^{ijkls})^2 \nonumber \\
&+\sqrt{\frac{(\tilde K^i)^2}{l_M^6}+\frac{R_s^2 (\tilde K^s)^2}{l_M^6}
+\frac{(\tilde K^{ijkl})^2}{4!~l_M^{12}}+\frac{R_s^2 (\tilde K^{ijks})^2}
{3!~l_M^{12}}}
\end{align}
where the shifted charges include the effect of the moduli through
\begin{eqnarray}
\tilde m_s=& m_s + \frac{1}{2}B_{ij} m^{ij} + 
   \frac{1}{8}B_{ij} B_{kl} m^{ijkls}\ ,\quad
&\tilde m_i=m_i +A_i m^{s}+B_{ij} m^{js} +\frac{1}{2}\C_{ijk} m^{jk}\nonumber\\
\tilde m^{is}=&m^{is} +A_i m^{ij}+ \frac{1}{3!}\C_{jkl} m^{ijkls}\ ,\quad
&\tilde m^{ij}=m^{ij}+\frac{1}{2} B_{kl} m^{ijkls}\nonumber\\
\tilde m^{ijks}=&m^{ijks} +A_l m^{ijkls}\ ,\quad
&\tilde m^{ijkls}=m^{ijkls}\\
\tilde K^s=&K^s+A_i K^i+\frac{1}{3!}\C_{ijk} K^{ijks} \ ,\quad
&\tilde K^i=K^i+\frac{1}{2} B_{jk} K^{ijks}+\frac{1}{3!}\C_{jkl} 
   K^{ijkl} \nonumber\\
\tilde K^{ijks}=&K^{ijks}+A_l K^{ijkl}\ ,\quad
&\tilde K^{ijkl}=K^{ijkl}\nonumber
\end{eqnarray}
The composite charges $K$ of the string multiplet
are as usual expressed for 1/4-BPS states
in terms of those in the particle multiplet
through
\begin{equation}
K^s=m_i m^{is}\ ,\quad K^i=m_j m^{ij} + m_s m^{is}\ ,\quad
K^{ijks}=m_l m^{ijkls}\ , \quad K^{ijkl}=m_s m^{ijkls}
\end{equation}
and vanish for half-BPS states.
In the Sen-Seiberg limit, the only terms remaining are 
$(\tilde m^{s})^2$ at leading order, $(\tilde m^{ij})^2$ and 
$(\tilde m^{is})^2$ at next-to-leading order, together with the first and
third term under the square root. We thus obtain, in string variables,
\begin{align}
\label{hme}
\begin{split}
{\cal M}=&\frac{|\tilde m_s|}{g_s l_s} +
\frac{g_s l_s}{2|\tilde m_s|} (\tilde m_i)^2
+\frac{1}{2\cdot 2 g_s l_s^5|\tilde m_s|}{(\tilde m^{ij})^2}
+\frac{g_s}{2g_s l_s^3|\tilde m_s|}{(\tilde m^{is})^2}\\
&+\frac{1}{2l_s^3 |\tilde m_s|}
\sqrt{(\tilde K^i)^2+\frac{1}{4!~g_s^2 l_s^6} (\tilde K^{ijkl})^2}
\end{split}
\end{align}

On the other hand, the energy spectrum for NCSYM on a torus $T^d$ was derived
in full rigour in \cite{Konechny:1999rz} for up to $d=4$ by computing
the central charges in the superalgebra. For $d=3$, it is possible to introduce
a topological term $\lambda^{ijk} \Tr F_{0i}F_{jk}$ in the action which
only affects charge quantization. One then finds
\begin{align} \label{d=3}
E=& \frac{g_{YM}^{2}}{2V\dim E}
\left(n^{i} + \Theta^{ij}m_{j}+ \lambda^{i}\dim E + 
\lambda^{ijk}q_{jk}\right)^2
\nonumber \\
& + \frac{V}{4g_{YM}^{2}\dim E} \left(q_{ij} + \dim E~ \Phi_{ij}\right)^2
 + \frac{1}{\dim E} \sqrt{ (k_i)^2}
\end{align}
where $k_i = (m_{i}\dim E - q_{ij}(n^{j} + \Theta^{jk}m_{k}))$
and the contractions are performed with respect to the metric $g_{ij}$
or its inverse $g^{ij}$.
For $\Phi_{ij}=0$, this formula is in complete agreement with the M-theory
side (\ref{hme})
under the standard map (\ref{mmap}), upon identifying 
\begin{equation}
\lambda^{ijk}=\C_{ijk}\ ,\quad
p=m_s\ ,\quad n^i=m_{i}\ ,\quad q_{ij}=m^{ij}\ ,\quad m_i=m^{is}\ ,
\end{equation}
and dropping the charge $K^{ijkl}$ which vanishes by antisymmetry.

Finally let us consider the case $d=4$. The dimension now involves
the second Chern class $r$ through
$\dim E=p + \frac{1}{2}\Theta^{ij} q_{ij} + \frac{1}{8} \epsilon_{ijkl}
\Theta^{ij}\Theta^{kl} r $
and the BPS spectrum takes the form
\begin{align}\label{d=4}
E=& \frac{g_{YM}^{2}}{2V\dim E}
\left[n^{i} + \lambda^{i}\dim E  + \lambda^{ijk}(q+*\Theta r)_{jk} + 
\Theta^{ij}m_{j} \right]^2 \nonumber \\
&+  \frac{V}{4g_{YM}^{2}\dim E}
\left[(q+*\Theta r)_{ij} + \dim E~ \Phi_{ij}\right]^2
+\frac{1}{\dim E}\sqrt{ (k_i)^{2} + \frac{1}{g^{4}_{YM}}(k^2)}\ .
\end{align}
Here $k_{i} = 
(m_{i}\dim E - (q + *\Theta r)_{ij}(n^{j} + \Theta^{jk}m_{k}) +
(*\lambda_{3})_{i}(ps-\epsilon^{jklm}q_{jk}q_{lm}/8))
$, $k=p r-\epsilon^{ijkl} q_{ij} q_{kl}/8 )^{2}$
 and $(*\lambda_{3})_{i}=\frac{1}{3!}\epsilon_{ijkl}\lambda^{jkl}$.
Again, for $\Phi_{ij}=0$ 
this is in perfect agreement with the M-theory mass formula up
to the rest term, upon identifying $r=\epsilon_{ijkl}m^{ijkls}/4!$.

\section{Magnetic background versus non-commutativity}

\noindent {\it The magnetic background $\phi$ as a compensator field}.
We now would like to offer a second resolution of the puzzle,
which yields a much deeper insight into the role of the modulus $\phi$.
Let us first focus on the first term in the NCSYM energy
(\ref{energy}), together with the rest term. We can rewrite the two
contributions in a more suggestive way as
\begin{equation}
\label{etot}
E_{rest}+E_{flux}=\frac{1}{l_s} \sqrt{\frac{V}{g_{YM}^4 l_s^4}} \left(
\frac{|p-q \theta|}{\sqrt{t}} 
+\frac{t^{3/2} \left[q-\phi (p-q\theta) \right]^2}{2 (p-q\theta)}
\right)
\end{equation}
where we defined $t=l_s^2/V$. This expression is as we argued
exactly invariant under the Morita transformation rules (\ref{tdualexo}),
or in terms of the present variables
\begin{equation}
\theta\to \frac{a\theta+b}{c\theta+d}\ ,\quad
\phi\to(c\theta+ d)^2 \phi - c (c\theta+ d)\ ,\quad
t\to t/(c\theta+d)^2 
\end{equation}
leaving $l_s$ and $V/g_{YM}^4$ invariant. In fact, the expression (\ref{etot})
arises as the two leading terms in the small $t$ 
expansion of the square root of an hypothetical ``relativistic'' 
 generalisation of (\ref{etot}),
\begin{equation}
\label{ncbi}
\tilde{\cal M}^2=\frac{V}{g_{YM}^4 l_s^6} \frac{\left[m-n\theta\right]^2
+ t^2 \left[n-\phi (m-n\theta) \right]^2  }{t} 
\end{equation}
where each term is separately invariant under the NCSYM duality
(\ref{tdualexo}). We claim that (\ref{ncbi}) is (part of) the mass formula of
Born-Infeld theory on a non-commutative torus, and will return to this claim
shortly.

The $Sl(2,\Zint)$ invariance of (\ref{ncbi}) seems to be realized in 
a quite different way from that of the standard $Sl(2,\Zint)$ invariant mass
formula 
\begin{equation}
\label{ncbiu}
{\cal M}^2=\frac{V}{g_{YM}^4 l_s^6} \frac{|m-nT|^2}{T_2} 
\end{equation}
which appears as the first term of (\ref{mmass}). In order to see
the relation between the two, we note that these mass formulae
can be recast in the form 
\begin{equation}
\label{meem}
{\cal M}^2= \begin{pmatrix} m & n \end{pmatrix}  
\cdot e^t e \cdot
\begin{pmatrix} m \\ n \end{pmatrix} 
\end{equation}
where $e$ is an element of $Sl(2,\Real)$ in both cases:
\begin{equation}
e=\frac{1}{\sqrt{T_2}}
\begin{pmatrix} 1 & -T_1  \\ 0 & T_2 \end{pmatrix} \ ,\qquad
\tilde e=\frac{1}{\sqrt{t}}
\begin{pmatrix} 1 & -\theta  \\ t\phi & (1-\theta\phi)t \end{pmatrix}
\end{equation}
Now, in contrast to $e$, 
$\tilde e$ is not an upper triangular matrix, but instead a 
$LU$ product
\begin{equation}
\tilde e=\frac{1}{\sqrt{t}}
\begin{pmatrix} 1 & 0  \\ t\phi & 1\end{pmatrix} \cdot
\begin{pmatrix} 1 & -\theta  \\ 0 & t \end{pmatrix}
\end{equation}
This can be brought into an upper triangular form through a $SO(2)$
rotation from the left, so that $\tilde e$ can be rewritten as
\begin{equation}
\tilde e= \Omega(\alpha)\cdot \frac{1}{\sqrt{t(1+t^2 \phi^2})}
\begin{pmatrix} 1+t^2\phi^2 & -(1+t^2\phi^2)\theta+t^2\phi\\
0 & t \end{pmatrix}
\end{equation}
with
\begin{equation}
\Omega(\alpha)=
\begin{pmatrix} \cos\alpha & -\sin\alpha \\ \sin\alpha & \cos\alpha
      \end{pmatrix}
\ ,\quad
\sin\alpha=\frac{t\phi}{\sqrt{1+t^2\phi^2}}\ ,\quad
\cos\alpha=\frac{1}{\sqrt{1+t^2\phi^2}}\ .
\end{equation}
The rotation drops from the mass formula (\ref{meem}),
so that the two mass formulae (\ref{ncbi}) and (\ref{ncbiu})
agree upon identifying the two upper triangular $Sl(2,\Real)$ elements
$e$ and $\Omega(\alpha)^{-1}\tilde e$, {\it i.e.}
\begin{equation}
\label{map}
T=\left(\theta-\frac{t^2 \phi}{1+t^2\phi^2}\right) + i \frac{t}{1+t^2\phi^2}
= \theta + i t \cos \alpha e^{i\alpha}
\end{equation}
It is easy to check that under this change of variables, the Morita
equivalence transformation rules (\ref{tdualexo}) imply the usual
T-duality ones (\ref{tdualsta}).
The interpretation of the modulus $\phi$ is thus quite clear: it,
or rather its associated angle $\alpha=\arctan(t\phi)$,
parametrizes the $SO(2)$ maximal compact subgroup of $Sl(2,\Real)$,
which leaves the mass formula (\ref{meem}) invariant. This implies
that the mass formula (\ref{ncbi})
is invariant under translations of $\alpha$, 
\begin{gather}
\label{so2}
\alpha\to\alpha+\beta\ ,\quad
t\to t \frac{\cos^2\alpha}{\cos^2(\alpha+\beta)}\ ,\quad
\theta \to \theta + t \frac{\cos(\alpha) \sin(\beta)}{\cos(\alpha+\beta)}
\end{gather}
which identifies various $\theta,\phi,t$ three-ples\footnote{The 
existence of such 
a symmetry was envisaged in \cite{Hofman:1998iy}, 
where it was rejected as unlikely.}. 
In particular, it is always possible to choose $\phi=0$, upon 
redefining $\theta$ and $t$, or $\theta=0$, upon redefining
$\phi$ and $t$. At $\phi=0$, the B-field is identified with the
non-commutative deformation parameter $\theta$, and the metric 
on the NCSYM side is simply the inverse of the M-theory metric,
so that the gauge theory lives on the reciprocal torus, albeit
non-commutative. At $\theta=0$
instead, the map (\ref{map}) yields $T=-1/(\phi+i/t)$, and recalling
that $t=l_s^2/V$, we see that the gauge theory lives on 
the {\it commutative} torus which is T-dual to the M-theory torus.
We should emphasize that the
symmetry (\ref{so2}) holds only for the Born-Infeld mass formula (\ref{ncbi}),
and not for its non-relativistic limit (\ref{etot}), and therefore
only Born-Infeld theory should allow for this trade-off between
non-commutativity and background magnetic field.

\vskip 3mm
\noindent {\it Born-Infeld theory on a non-commutative two-torus.}
We can now include the effect of the other charges $n^i,m^i$ and $k_i$
into a relativistic generalisation of the NCSYM energy formula 
(\ref{energy}),
\begin{align}
\label{ncbi2}
\tilde{\cal M}^2=&\left( \frac{V}{g_{YM}^2 l_s^4} \right)^2 (\tilde p)^2
+ (\tilde n^i)^2 
+ \left( \frac{1}{g_{YM}^2 l_s} \right)^2 (\tilde q)^2 
+ \sqrt{\left(\frac{V}{g_{YM}^2 l_s^4} \right)^2 (k_i)^2 }
\end{align}
where the shifted charges incorporate the effect of the deformation
parameter $\theta$, the magnetic background $\phi$ and the topological
angles $\lambda^i$ through 
\begin{subequations}
\label{shiftc2}
\begin{eqnarray}
\tilde p&=& p  -\theta q\\
\tilde q&=& q  + \phi (p-\theta q)\\
\tilde n^i&=&n^i -\theta m^i +\lambda^i (p-\theta q)\\
\tilde m^i&=& m^i  - \phi [n^i-\theta m^i +\lambda^i 
(q  + \phi (p-\theta q)] + \lambda^i (p-\theta q)
\end{eqnarray}
\end{subequations}
This mass formula reduces to (\ref{energy}) in the
Sen-Seiberg non-relativistic limit $l_s\to 0$ holding
$V/g_{YM}^2$ and $g_{ij}/l_s^2$ fixed,
and to the usual Born-Infeld energy formula when $\theta=0$,
and coincides with the one proposed in
\cite{Hofman:1998iv} in the special case $\phi=0$.
We claim that this is the BPS energy formula for supersymmetric
Born-Infeld theory on a non-commutative torus. This theory
may be tentatively defined by the action functional
\begin{equation}
S=\frac{V}{2g_{YM}^2} \Tr  \sqrt{\det\left[g_{ij}+l_s^2 (F_{ij}+
\Phi_{ij}\cdot 1)\right]} + i \lambda^i \Tr F_{0i}\ ,
\end{equation}
but the determinant would need to be properly defined.
The resulting theory should exhibit the same Morita equivalence like its
well-defined NCSYM cousin. The invariance of the BPS mass formula 
under the $SO(2)$ symmetry
(\ref{so2}) hints to a dynamical symmetry of this putative theory,
which would allow to trade an Abelian magnetic background ($\phi\neq 0$)
for a non-commutativity of the base-space ($\theta\neq 0$).
Unfortunately, we are not able to prove this assertion at that stage. 
Given that
non-Abelian Yang-Mills on a commutative torus is Morita equivalent
to NCSYM at a rational value of the deformation parameter $\theta$,
one may hope to use this symmetry to shed light on non-Abelian Born-Infeld
on a commutative torus by relating it to its Abelian version
on a commutative torus with magnetic background.

\vskip 3mm
\noindent{\it Generalization to higher-dimensional tori.} 
We now would like to generalize our considerations to 
higher-dimensional tori. The standard T-duality invariant
mass formula in the vector representation of $SO(d,d,\Zint)$
takes the form
\begin{equation}
\label{esta}
{\cal M}^2=\begin{pmatrix} m_i & m^i \end{pmatrix}\cdot
e^t e\cdot \begin{pmatrix} m_i \\ m^i \end{pmatrix}\ ,\quad
e= \begin{pmatrix}(V^{-1})^t & \\ & V \end{pmatrix}\cdot
\begin{pmatrix} 1 & B \\ & 1 \end{pmatrix}\cdot
\end{equation}
where $e$ is the general form of a $SO(d,d,\Real)$ element
up to left action of the maximal compact subgroup $SO(d)\times SO(d)$.
$V$ is the vielbein of the metric $G=V^t V$, and can be chosen
in an upper triangular form. The action of $SO(d,d,\Zint)$
then takes the standard form
\begin{gather}
\label{tduald}
G+B\to  (\A(G+B)+\B)(\C(G+B)+\D)^{-1}\ ,\quad
\begin{pmatrix} m_i \\ m^i \end{pmatrix} \to 
\begin{pmatrix}\A & \B\\ \C & \D \end{pmatrix}
\begin{pmatrix} m_i \\ m^i \end{pmatrix}\ ,\quad
\end{gather}
where $\A,\B,\C,\D$ are integer-valued matrices parametrizing
an $SO(d,d,\Zint)$ element,
\begin{equation}
\A^t\C=-\C^t\A\ ,\quad \B^t\D=-\D^t\B\ ,\quad \A^t\D+\C^t\B=1
\end{equation}

The experience with the $d=2$ case as well as the way $\Phi$ enters
in the general energy formula (\ref{d=4}) suggests that the appropriate element
of $SO(d,d)$ parametrized by the moduli $g,\Theta,\Phi$ is instead
\begin{equation}
\label{etilded}
\tilde e=
\begin{pmatrix} (v^{-1})^t  &   \\  & v \end{pmatrix} \cdot
\begin{pmatrix} 1 & \\ \Phi & 1 \end{pmatrix} \cdot
\begin{pmatrix} 1 & \Theta  \\  & 1 \end{pmatrix} \cdot
\end{equation}
where $v$ is the vielbein of the inverse metric $g^{-1}=v^tv$.
This mass formula is invariant under the Morita equivalence
transformation rule,
\begin{subequations}
\label{moritad}
\begin{gather}
\Theta\to  (\A\Theta+\B)(\C\Theta+\D)^{-1}\ ,\quad
\Phi\to(\C\Theta+ \D)\Phi(\C\Theta+ \D)^t + \C (\C\Theta+ \D)^t\ ,\quad \\
g \to (\C\Theta+\D)g (\C\Theta+\D)^t\ ,
g_{YM}^2\to\sqrt{|\det(\C\Theta+\D)|}g_{YM}^2\ ,\\
\begin{pmatrix} n^i \\ n_i \end{pmatrix} \to 
\begin{pmatrix}\A & \B\\\C & \D \end{pmatrix}
\begin{pmatrix} n^i \\ n_i \end{pmatrix}\ ,\quad
\end{gather}
\end{subequations}
We now want to find a mapping between $(g,\Theta,\Phi)$
and $(G,B)$ which translates (\ref{moritad}) into (\ref{tduald}).
For this, we rotate the element $\tilde e$ 
into a block upper triangular form through
a $SO(d)$ action from the left,
\begin{equation}
\label{erot}
\tilde e= 
\begin{pmatrix} \frac{1+\Omega}{2} & \frac{1-\Omega}{2} \\ 
\frac{1-\Omega}{2} & \frac{1+\Omega }{2}
\end{pmatrix} \cdot
\begin{pmatrix} (1+v\Phi v^t)(v^{-1})^{t}&\\ & (1-v\Phi v^t)^{-1}v
 \end{pmatrix} \cdot
\begin{pmatrix} 1 & \Theta-(g^{-1}-\Phi g \Phi)^{-1} \Phi g
\\ & 1 \end{pmatrix}
\end{equation}
where the rotation parameter is given by the orthogonal matrix
\begin{equation}
\Omega=(1-v\Phi v^t)(1+v\Phi v^t)^{-1}\ .
\end{equation}
The $SO(d)$ rotation drops from the mass formula, so that we can 
identify (\ref{erot}) to (\ref{esta}),
\begin{equation}
G=(g-\Phi g^{-1} \Phi)^{-1}\ ,\quad 
B=\Theta-(g-\Phi g^{-1} \Phi)^{-1}\Phi g^{-1}
\end{equation}
or, combining the symmetric and antisymmetric matrices,
\begin{equation}
\label{mapd}
G+B=\Theta+(g+\Phi)^{-1}\ .
\end{equation}
This can be equivalently expressed in terms of the variable $\Omega$,
\begin{equation}
G=v^t \frac{(\Omega+1)^2}{4\Omega} v \ ,\quad 
B= \Theta - v^t \frac{\Omega^2-1}{4\Omega}v\ .
\end{equation}
$\Phi$ thus appears as a compensator field
for the $SO(d)$ compact subgroup of the continuous T-duality group 
$SO(d,d,\Real)$\footnote{The invariance under the other $SO(d)$ compact 
subgroup is a consequence of the Lorentz invariance of the theory.}.
Under this change of variables, the Morita equivalence (\ref{moritad})
becomes the standard T-duality map (\ref{tduald}).
For $\Phi=0$, the non-commutative deformation parameter $\Theta$
is equated to the M-theory B-field, while the metric is the
inverse of the M-theory metric, so that the gauge theory
lives in the deformed momentum space. For $\Theta=0$ instead, 
the relation (\ref{mapd}) becomes $G+B=(g+\Phi)^{-1}$,
so that the gauge theory lives on the commutative T-dual torus.
For $d=2$, we recover the previous result
(\ref{map}) upon setting $v=\sqrt{t},\ \Theta=i\theta,\ \Phi=i\phi,\ 
\Omega=e^{-i\alpha}$.

As in the $d=2$ case, there is a continuous symmetry $SO(d)$
which rotates $\tilde e$ and implies an action on the moduli
$v,\Theta,\Phi$. The exponentiated action is cumbersome,
but the infinitesimal transformation is easily derived,
\begin{equation}
\label{sod}
dv=(\Omega+1)^{-2}\Omega d\Omega\ v\ ,\quad
d\Theta=\frac{1}{4}v^t \left[
\frac{\Omega-1}{\Omega+1} d\Omega + d\Omega^t \frac{\Omega-1}{\Omega+1} 
\right] v\ .
\end{equation}
by requiring the left-hand side of the map (\ref{mapd}) to stay
invariant. Again, we stress that this symmetry is a property of the
relativistic dispersion relation, and not of its Galilean limit.

\vskip 3mm
\noindent {\it Born-Infeld theory on a non-commutative $d$-torus.}
In the same way as for $d=2$, we can now use these results to derive
relativistic generalizations of the BPS energy formulae
(\ref{d=3},\ref{d=4}), which reduce to them in the decoupling limit $l_s\to 0$
and reproduce the $d=2$ result (\ref{ncbi}) in the decompactification
limit. One way to proceed is to start from the M-theory mass formula
(\ref{genmass}), translate it in gauge theory variables through the
relations (\ref{mapd}) at $\Phi=0$, and switch on the modulus $\Phi$ 
according to the mass formula (\ref{etilded}) translated into the appropriate
representation of $SO(d,d)$. Restricting ourselves
to the case $d\leq 4$ , we obtain
\begin{align}
\label{genmassbi}
\tilde{\cal M}^2=&\left( \frac{V}{g_{YM}^2 l_s^4} \right)^2 (\tilde p)^2
+ \frac{1}{l_s^4}(\tilde m^i)^2 
+ (\tilde m_i)^2 
+ \left( \frac{V}{g_{YM}^2 l_s^2} \right)^2 \frac{(\tilde m_{ij})^2 }{2!}
+ \left( \frac{V}{g_{YM}^2 } \right)^2 \frac{(\tilde m_{ijkl})^2}{4!} 
\nonumber\\
&+ \sqrt{\left(\frac{V}{g_{YM}^2 l_s^4} \right)^2 \tilde (k_i)^2 
+ \frac{1}{l_s^4}(\tilde k)^2 
+\left(\frac{V}{g_{YM}^2 l_s^2} \right)^2 \frac{\tilde (k_{ijk})^2 }{3!}
+\left(\frac{V l_s}{g_{YM}^2} \right)^4 \frac{\tilde (k_{ijkl})^2}{4!} 
}
\end{align}
where the shifted charges incorporate the effect of the moduli:
\begin{subequations}
\begin{gather}
\left\{ \begin{array}{ccc}
\tilde p&=& p + \frac{1}{2}\Theta^{ij} m^{ij} + 
   \frac{1}{8}\Theta^{ij} \Theta^{kl} m_{ijkl}\ \\
\tilde m_{ij}&=&m_{ij}+\frac{1}{2} \Theta^{kl} m_{ijkl}
 +\frac{1}{2} \Phi_{ij} \tilde p \\
\tilde m_{ijkl}&=& m_{ijkl}-\frac{1}{2} \Phi_{[ij} \tilde m_{kl]}
+\frac{1}{8} \Phi_{[ij}\Phi_{kl]}~ \tilde p \end{array}
\right.\\
\left\{ \begin{array}{ccc}
\tilde m^i&=&m^i +\Theta^{ij} m_{j} +\lambda^{ijk} m_{jk} +\lambda^i \tilde p \\
\tilde m_{i}&=&m_{i} +\Phi_{ij} m^j+
   \frac{1}{3!}\lambda^{jkl} m_{ijkl}+\lambda^j \tilde m_{ij}\  \end{array}
\right.\\
\left\{ \begin{array}{ccc}
\tilde k_i&=&k_i+\Theta^{jk} k_{ijk}+\frac{1}{3!}\lambda^{jkl} k_{ijkl} \\
\tilde k_{ijk}&=&k_{ijk} +\frac{1}{2}\Phi_{[ij} \tilde k_{k]} +\lambda^l 
\tilde k_{ijkl} \end{array}
\right.\\
\tilde k  =k+\frac{1}{3!}\lambda^{ijk} k_{ijk} +A^i \tilde k_i\ ,
\quad \tilde k_{ijkl}=k_{ijkl}
\end{gather}
\end{subequations}
Here we gathered the charges according to their representation under
$SO(4,4)$: conjugate spinor ($p,m_{ij},m_{ijkl}$), vector ($m^i,m_i$), 
spinor ($k_i,k_{ijk}$), singlets ($k,k_{ijkl}$) respectively. 
For 1/4-BPS states, the momentum multiplet charges 
are related to the flux multiplet charges through 
\begin{equation}
k=m_i m^{i}\ ,\quad k_i=m^j m_{ij} + m~ m_{i}\ ,\quad
k_{ijk}=m^l m_{ijkl}\ , \quad k_{ijkl}=m~m_{ijkl}
\end{equation}
and vanish for half-BPS states.
The energy formula (\ref{genmassbi}) is invariant under
the Morita equivalence transformation (\ref{moritad})
supplemented by the laws
\begin{subequations}
\begin{eqnarray}
\lambda^{ijk} \to \sqrt{|\det(\K)|} \K^{i}_{l}\K^{j}_{m}\K^{k}_{n} \lambda^{lmn}
\ ,\quad\\
\lambda^{i}\to\sqrt{|\det(\K)|} \K^{i}_{j} \lambda^{j} -[\C \K^{-1}]^{j}_{k}
\sqrt{|\det(\K)|} \K^{i}_{l}\K^{j}_{m}\K^{k}_{n} \lambda^{lmn}
\end{eqnarray}
\end{subequations}
where $\K=((\C\Theta+\D)^t)^{-1}$,
as appropriate for a spinor representation of $SO(d,d)$,
\cite{Schwarz:1998qj,Brace:1998xz,Konechny:1999rz}. Besides, in contrast to
its non-relativistic limit, it is by construction invariant
under the full U-duality group, $SO(5,5,\Zint)$ in the
maximal case $d=4$ that we considered, which in the context of 
M(atrix) theory was named extended U-duality
\cite{Blau:1997du,Obers:1997kk,Hull:1997jc}. It is also invariant
under the $SO(d)$ symmetry (\ref{sod}) which allows to 
trade the deformation parameter $\Theta$ for the
magnetic background $\Phi$. For $\Phi=0$, 
it agrees with the formula constructed in \cite{Hofman:1998iv}.
Equation (\ref{genmassbi}) should therefore give the 
energy of 1/4-BPS states of supersymmetric Born-Infeld on a 
non-commutative torus $T^d$ with arbitrary magnetic background
$\Phi$, once this theory receives an appropriate definition. 

\acknowledgments
The authors are grateful to the organizers of
the workshop ``D-branes, vector bundles and bound states'' at IHES, 
Bures-sur-Yvette, June 1999
for providing a stimulating atmosphere, and to IHES and CERN Theory Division
for their kind hospitality during part of this work. We are indebted to
M. Douglas and E. Rabinovici for very useful discussions.

\vskip .5cm
\noindent {\it Note added in proof.} After our work was released, a
related work appeared on the archive \cite{sw2}, discussing the
relevance of non-commutative gauge theories in the context of
D-branes. In particular, the equivalence of Born-Infeld gauge theories
on a non-commutative torus under continuous shifts of $\Theta$ was
proved und the field redefinition  explicitly constructed. This is
closely related to the $SO(d)$ symmetry uncovered in the
present work on the base of consideration of the BPS spectrum. 
The relation \eqref{mapd} between 
the gauge theory and space-time moduli agrees with Eq. (4.3) in
\cite{sw2}, upon identifying our gauge theory metric $g$ with their
open string metric $G$, and our target-space metric and B-field $G+B$ with the
T-dual of their closed string metric and B-field $g+B$, as it should since our
discussion was from the point of view of D0-branes instead of D$d$-branes.


\providecommand{\href}[2]{#2}\begingroup\raggedright\endgroup

\end{document}